\documentclass[twocolumn,aps,showpacs]{revtex4}
\usepackage{amsmath}
\usepackage{amssymb}
\usepackage{epsf}
\begin{document}

\title{Influence of trigonal warping on interference effects in bilayer graphene}
\author{K. Kechedzhi$^{1}$, Vladimir I. Fal'ko$^{1}$, E. McCann$^{1}$, and
B.L. Altshuler$^{1,2}$}
\affiliation{$^{1}$Department of Physics, Lancaster University, Lancaster, LA1 4YB,~UK \\
$^{2}$Physics Department, Columbia University, 538 West 120th Street, New
York, NY 10027}

\begin{abstract}
Bilayer graphene (two coupled graphitic monolayers arranged
according to Bernal stacking) is a two-dimensional gapless
semiconductor with a peculiar electronic spectrum different from
the Dirac spectrum in the monolayer material. In particular, the
electronic Fermi line in each of its valleys has a strong
$\mathbf{p}\rightarrow -\mathbf{p}$ asymmetry due to trigonal
warping, which suppresses the weak localization effect. We show
that weak localization in bilayer graphene may be present only in
devices with pronounced intervalley scattering, and we evaluate
the corresponding magnetoresistance.
\end{abstract}


\maketitle

Weak localization (WL), which universally occurs in a broad range
of disordered conductors \cite{WL,WLso} is caused by the
constructive interference of electron waves circling the same
closed path in opposite directions. It is sensitive to the
presence of an external magnetic field as manifested in a negative
magnetoresistance (MR) at low temperatures.

Usually qualitative features of WL do not depend on the detail of
the electronic band structure and crystalline symmetry, with the
exception of spin-orbit coupling \cite{WLso,dresselhaus92}. In
gapful multi-valley semiconductors only the size of the WL effect
may depend on the number of valleys and the strength of
intervalley scattering \cite{fuku80,bishop80,prasad95}. The
low-field MR, $\Delta \rho (B)\equiv \rho (B)-\rho (0)$, in a two
dimensional electron gas or a thin metallic film
\cite{socomment,WL,WLso,fuku80} in the absence of spin-orbit
coupling is characterized by
\begin{eqnarray*}
\Delta \rho (B) &=& -\frac{s_{\theta} e^{2}\rho ^{2}}{2\pi
h}F\left( \frac{B}{B_{\varphi }} \right) ,\;B_{\varphi
}=\frac{\hbar c}{4De}\tau _{\varphi }^{-1}.  \label{summary}
\end{eqnarray*}
Here $F(z)=\ln z+\psi (\frac{1}{2}+\frac{1}{z})$, $\psi(z)$ is the
digamma function, $\tau_{\varphi}$ is the coherence time, $D$ is
the diffusion coefficient, and the integer factor $s_{\theta}$
depends on whether or not states in $n_v$ valleys are mixed by
disorder. This factor is controlled by the ratio $\theta =
\tau_{\textrm{i}} / \tau_{\varphi}$ between the intervalley
scattering time $\tau_{\textrm{i}}$ and the coherence time
$\tau_{\varphi}$. In materials such as Mg, ZnO, Si, Ge (listed in
Table~I) where each of the Fermi surface pockets is
$\mathbf{p}\rightarrow -\mathbf{p}$\ symmetric, intervalley
scattering reduces the size of the WL MR from that described by
$s_{\infty} = 2n_{v}$ when $\theta = \tau_{\textrm{i}} /
\tau_{\varphi} \gg 1$ to $s_0 = 2$ for $\theta \ll 1$.

A more interesting scenario develops in a multi-valley semimetal,
where the localization properties can be influenced by the absence
of $\mathbf{p} \rightarrow - \mathbf{p}$\ symmetry of the
electronic dispersion within a single valley. Graphene
\cite{DD,QHEbi,McCann} in a graphene-based transistor
\cite{NovScience,Novoselov,Zhang} represents an example of such a
system. In this Letter, we demonstrate how the asymmetry in the
shape of the Fermi surface in each of its two valleys determines
the observable WL behavior. In contrast with conventional
materials (see Table~I), WL MR in bilayer graphene is increased by
the intervalley scattering, from $s_{\infty} = 0$ at $\theta
\rightarrow \infty$ to $s_0=2$ at $\theta \ll 1$.

\begin{table}[b]
\caption{WL factor $s_{\theta}$ in conductors with a multi-valley
conduction band and negligible spin-orbit coupling. The factor
$s_{\theta}$ is specified for two limiting cases, no inter-valley
scattering $\theta = \tau_{\textrm{i}} / \tau_{\varphi}
\rightarrow \infty$, and for fast inter-valley scattering $\theta
\rightarrow 0$.}
\begin{tabular}{clcc}
\hline\hline \hspace{0.2cm}$n_{v}$\hspace{0.2cm} &
& \hspace{0.2cm}$s_{\infty }$\hspace{0.2cm} & \hspace{0.2cm}$s_{0}$\hspace{0.2cm}\\
\hline $1$
& \multicolumn{1}{l}{Mg films \cite{bergmann84}, ZnO wells \cite{goldenblum99}}
& $2$ & - \\
$2,6$ & \multicolumn{1}{l}{Si MOSFETs \cite{fuku80,bishop80}} &
$2n_{v}$ & $2$ \\
$2$ & \multicolumn{1}{l}{Si/SiGe wells \cite{prasad95}} & $4$ & $2$ \\
$2$ & \multicolumn{1}{l}{graphene} & $0$ & $2$ \\
\hline\hline
\end{tabular}
\end{table}

Bilayer graphene consists of two coupled graphitic monolayers
arranged according to Bernal stacking, see \cite{McCann} for
details of the lattice configuration. Its unit cell contains sites
$A,B$  and  $\tilde{A}$, $\tilde{B}$. Sites $A,B$ and
$\tilde{A},\tilde{B}$ belong to the honeycomb lattice in the
bottom and top layers, respectively, with sites $B$ being exactly
below $\tilde{A}$. The hexagonal Brillouin zone of the bilayer has
two inequivalent corners $\mathbf{K}_{+}$ and $\mathbf{K}_{-}$
\cite{kpoints}. The four branches of its electronic spectrum
\cite{ohta06} form one pair split by about $\pm \gamma _{1}$ (the
interlayer coupling) and two low-energy branches (formed by states
based upon sublattices $A$ and $\tilde{B}$) which are degenerate
at $\mathbf{K}_{+}$ and $\mathbf{K}_{-}$. The low-energy branch
can be described \cite{McCann} using a Hamiltonian,
\begin{eqnarray}
&&\!\!\!\!\!\!\!\!\!\!{\hat{H}}_{2L}=-\frac{1}{2m}\left[ \left(
{p}_{x}^{2}-{p}_{y}^{2}\right) \sigma _{x}+2{p}_{x}{p}_{y}\sigma
_{y}\right]
+{\hat{h}}_{\mathrm{w}},  \label{h2} \\
&&{\hat{h}}_{\mathrm{w}}=v_{3}\Pi _{z}\left( p_{x}\sigma
_{x}-p_{y}\sigma _{y}\right), \notag
\end{eqnarray}
which acts in the space of four-component wave functions $\Phi
=[\phi _{ \mathbf{K}_{+},A},\phi _{\mathbf{K}_{+},\tilde{B}},\phi
_{\mathbf{K}_{-}, \tilde{B}},\phi _{\mathbf{K}_{-},A}]$. Here,
$\phi _{\xi ,\alpha }$ is an electron amplitude on the sublattice
$\alpha =A,\tilde{B}$ in the valley $\xi = \mathbf{K_{+}},
\mathbf{K_{-}}$, $\sigma_{x,y,z}$ and $\Pi_{x,y,z}$ are Pauli
matrices acting in sublattice and valley space, respectively
\cite{kpoints}. In slightly doped graphene, two disconnected Fermi
lines surround the corners of the Brillouin zone.

\begin{figure}[b]
\centerline{\epsfxsize=1\hsize\epsffile{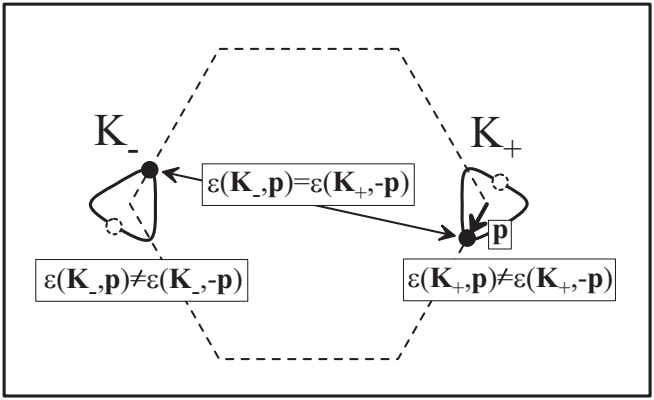}}
\caption{Fermi lines (solid lines) in the vicinity of two
inequivalent valleys $\mathbf{K}_{+}$ and $\mathbf{K}_{-}$ of the
hexagonal Brillouin zone (dashed line). Trigonal warping produces
asymmetry of the dispersion at each valley $\protect\epsilon
(\mathbf{K}_{\pm },\mathbf{p})\neq \protect \epsilon
(\mathbf{K}_{\pm },-\mathbf{p})$, where momentum $\mathbf{p}$ is
determined with respect to the center of the valley, but the
effects of warping in the valleys have opposite signs,
$\protect\epsilon (\mathbf{K}_{\pm },\mathbf{p} )=\protect\epsilon
(\mathbf{K}_{\mp },-\mathbf{p})$. The bilayer lattice
configuration is described in detail in \cite{McCann}.}
\label{fig:1}
\end{figure}

The first term in Eq.~(\ref{h2}) is the leading contribution in
the nearest neighbor approximation of the bilayer tight binding
model \cite{McCann}. It includes intralayer hopping
$A\leftrightarrows B$ and $\tilde{A} \leftrightarrows \tilde{B}$
(that leads to the dispersion $\epsilon =\pm pv$ near
$\mathbf{K_{\pm }}$ in a monolayer) and interlayer $\tilde{A}
\leftrightarrows B$ hopping, and it yields a parabolic spectrum
$\epsilon =\pm p^{2}/2m$ with $m=\gamma _{1}/2v^{2}$. It has been
noticed \cite{McCann} that quasiparticles described by this term
are chiral \cite{notegf}: their plane wave states are eigenstates
of an operator $\mathbf{\sigma n}_{2}$ with $\mathbf{\sigma
n}_{2}=-1$ for electrons in the conduction band, and
$\mathbf{\sigma n}_{2}=1$ for the valence band, where
$\mathbf{n}_{2}(\mathbf{p})=(\text{cos}(2\varphi
),\text{sin}(2\varphi ))$ for $\mathbf{p}=(p\text{cos}\varphi
,p\text{sin}\varphi )$. For an electron in the conduction band,
the plane wave state is
\begin{equation*}
\Phi _{\mathbf{K},\pm \mathbf{p}}=\pm \frac{e^{\pm
i\mathbf{px}/\hbar }}{\sqrt{2}}\left( e^{-i\varphi }|\uparrow
\rangle _{\mathbf{K},\pm \mathbf{p}}-e^{i\varphi }|\downarrow
\rangle _{\mathbf{K},\pm \mathbf{p}}\right) ,
\end{equation*}
where ${|\uparrow \rangle_{\mathbf{K}_{+},\mathbf{p}}}=[1,0,0,0]$,
${|\downarrow \rangle_{\mathbf{K}_{+},\mathbf{p}}}=[0,1,0,0]$ and
${|\uparrow \rangle_{\mathbf{K}_{-},\mathbf{p}}}=[0,0,1,0]$,
${|\uparrow \rangle_{\mathbf{K}_{-},\mathbf{p}}}=[0,0,0,1]$, and
the factors $e^{\pm i\varphi }$ take into account the chirality.

The term ${\hat{h}}_{\mathrm{w}}$ in Eq.~(\ref{h2}) originates
from a weak direct $A\leftrightarrows \tilde{B}$ interlayer
coupling. This one gives rise to $\mathbf{p}\rightarrow
-\mathbf{p}$\ asymmetry in the dispersion of electrons. It also
leads to a Lifshitz transition in the shape of the Fermi line of
the 2D electrons which takes place when $\epsilon_{\mathrm{F}}\sim
\epsilon_{\mathrm{L}} = {\textstyle\frac{1}{4}}
\gamma_{1}(v_{3}/v)^{2}$ thus $n_{e} > n_{\mathrm{L}} \approx
v_{3}^2\gamma_{1}^{2}/(2\pi\hbar^{2}v^{4}) \sim
10^{11}\textrm{cm}^{-2}$ (using $v_{3}/v\sim 0.1$, Ref.
\cite{McCann}). As long as $n_{e}< n_{\mathrm{L}}$, the Fermi line
in each valley is split into four pockets \cite{McCann}. For
$n_{e} > n_{\mathrm{L}}$, ${\hat{h}}_{\mathrm{w}}$ can be treated
as a perturbation leading to a trigonal deformation of a
single-connected Fermi line and asymmetry of the electron
dispersion inside each valley illustrated in Fig. 1: $\epsilon
(\mathbf{K}_{\pm },\mathbf{p})\neq \epsilon (\mathbf{K}_{\pm
},-\mathbf{p})$, though, due to time-reversal symmetry
\cite{t-rev}, $\epsilon (\mathbf{K}_{\pm },\mathbf{p})= \epsilon
(\mathbf{K}_{\mp },-\mathbf{p})$. Existing graphene structures are
strongly affected by charges trapped in the underlying substrate
or on its surface. Such disorder (as well as the inhomogeneity of
the electron density) obscures the intricate details of the
electronic Fermi surface at the lowest energies, $\epsilon \ll
\epsilon_{\mathrm{L}}$. However the interplay between the two
terms in ${\hat{H}}_{2L}$ resulting in intravalley asymmetry of
the electronic dispersion manifests itself in the WL behavior.

The WL correction to conductivity in disordered conductors is a
result of the interference of electrons propagating around closed
loops in opposite directions \cite{WL}. Such interference is
constructive in materials without spin-orbit coupling, since
electrons acquire exactly the same phase when travelling along two
time-reversed paths. It is usually described \cite{WL} in terms of
the particle-particle correlation function, Cooperon. To evaluate
WL in bilayer graphene, we use a Cooperon matrix $C_{\alpha \beta
\alpha ^{\prime }\beta ^{\prime }}^{\xi \mu \xi ^{\prime }\mu
^{\prime }}$ where subscripts label the sublattice state of
incoming $\alpha \beta $ and outgoing $\alpha ^{\prime }\beta
^{\prime }$ pairs of electrons and superscripts describe the
valley state of incoming $\xi \mu $ and outgoing $\xi ^{\prime
}\mu ^{\prime }$ pairs. Following the example of Cooperons for a
spin ${\textstyle\frac{1}{2}}$ \cite{WLso}, we classify Cooperons
as singlets and triplets in terms of sublattice and valley indices
$C_{S_{1}S_{2}}^{M_{1}M_{2}}$. For example, $M=0$ is a
"valley-singlet", $M=x,y,z$ are three "valley-triplet" components;
$S=0$ is a "sublattice-singlet" and $S=x,y,z$ are
"sublattice-triplet" components.

With regards to the sublattice composition of Cooperons in
disordered bilayer graphene, only modes $C_{z}^{M}$ (with $S=z$
\cite{holes}) are relevant. Indeed, a correlator $C \sim \Phi
_{\mathbf{K}_+,\mathbf{p}}\Phi _{\mathbf{K}_-,-\mathbf{p}}$
describing two plane waves, $\Phi _{\mathbf{K}_+,\mathbf{p}}$ and
$\Phi _{\mathbf{K}_-,-\mathbf{p}}$, propagating in opposite
directions along a ballistic segment of a closed trajectory
contains only sublattice-triplet terms,
\begin{eqnarray}
C &\sim& \left( {|\uparrow \rangle
}_{\mathbf{K}_+,\mathbf{p}}|\downarrow \rangle
_{\mathbf{K}_-,-\mathbf{p}}\! +\!|\downarrow \rangle
_{\mathbf{K}_+,\mathbf{p}}|\uparrow \rangle _{\mathbf{K}_-,-\mathbf{p}}\!\right)   \notag \\
&-&e^{2i\varphi }|\uparrow\rangle
_{\mathbf{K}_+,\mathbf{p}}|\uparrow \rangle
_{\mathbf{K}_-,-\mathbf{p}}\!-e^{-2i\varphi }|\downarrow \rangle
_{\mathbf{K}_+,\mathbf{p}}|\downarrow \rangle
_{\mathbf{K}_-,-\mathbf{p}},  \notag
\end{eqnarray}
and the terms corresponding to $C_{x,y}^{M}$ disappear after
averaging over the direction of momentum, $\langle e^{\pm
2i\varphi }\rangle _{\varphi }=0$.

With regards to the valley composition of relevant Cooperon modes,
the symmetry of the electronic dispersion within each valley plays
a pivotal role. For a conventional metal \cite{bergmann84} or
semiconductor \cite{fuku80,bishop80,prasad95,goldenblum99}, two
phases $\vartheta_{\circlearrowright }$ and
$\vartheta_{\circlearrowleft }$ acquired by an electron while
propagating in a clockwise and anti-clockwise direction along the
same loop are exactly equal, so that the interference enhances
backscattering and leads to WL \cite{WL}. Any closed trajectory is
a combination of ballistic intervals characterized by momenta $\pm
\mathbf{ p}_{j}$ (for two directions) and time of flight $t_{j}$.
The asymmetry of the electron dispersion in bilayer graphene, due
to ${\hat{h}}_{\mathrm{w}}$, generates a phase difference $\delta
\equiv \vartheta _{\circlearrowright }-\vartheta
_{\circlearrowleft }=\sum \delta _{j}$, where  $\delta _{j}\sim
[\epsilon (\mathbf{p_{j}})-\epsilon
(-\mathbf{p_{j}})]t_{j}=2{\hat{h}}_{\mathrm{w}}(\mathbf{p_{j}})t_{j}$.
Since $\delta _{j}$ are random uncorrelated, the mean square of
$\delta $ accumulated over the time interval $t = \sum_{j} t_{j}$
can be estimated as $\langle \delta ^{2}\rangle \sim \langle
(t_{j}{\hat{h}}_{\mathrm{w}}(\mathbf{p_{j}}))^{2}\rangle t/\tau$,
where $\tau$ is the transport time and $l \equiv v_F \tau$ is the
mean free path \cite{notegf}. This determines the relaxation rate,
\begin{equation}
\tau_{\mathrm{w}}^{-1} = \left\{
\begin{array}{cc}
{\textstyle\frac{1}{2\hbar^2}} \tau \langle
\mathrm{Tr}{\hat{h}}_{\mathrm{w}}^{2}(\mathbf{p})\rangle_{\varphi
} = \pi n_L l^2 \tau^{-1}, & \pi n_L l^2 < 1 \\
\!\!\!\!\!\!\!\!\!\!\!\!\!\!\!\!\!\!\!\!\!\!\!\!\!\!\!\!\!\!\!\!\!\!\!\!\!\!\!\!\!\!\!\!\!\!\!\!\!\!\!\!\!\!\!\!\!
\tau^{-1}, & \pi n_L l^2 > 1 \\
\end{array}\right., \label{tauw}
\end{equation}
which suppresses the intravalley Cooperons $C_{z}^{x}$,
$C_{z}^{y}$. We estimate that for the recently studied bilayers
\cite{QHEbi} with $n_{e}= 2.5 \times 10^{12}\textrm{cm}^{-2}$,
$\tau_{\mathrm{w}} \sim \tau$ and $l \sim 0.1 \mu\textrm{m}$. A
similar situation occurs in bilayer structures studied by
R.~Gorbachev {\em et al.} \cite{exeter}. Due to time-reversal
symmetry [$\epsilon (\mathbf{K}_{\pm },\mathbf{p})=\epsilon
(\mathbf{K}_{\mp },-\mathbf{p})$] \cite{t-rev}, signs of warping
in the valleys $\mathbf{K_{+}}$ and $\mathbf{K_{-}}$ are opposite,
so that it does not suppress the intervalley Cooperons $C_{z}^{0}$
and $C_{z}^{z}$.

The above qualitative analysis indicates that strong trigonal
warping should be associated with the suppression of WL in bilayer
graphene when electrons do not change their valley state upon
scattering. To give a quantitative description of this prediction
we used the diagrammatic technique and related the WL correction
to conductivity to the surviving Cooperon modes \cite{socomment},
\begin{equation}
\delta g=\frac{2e^{2}D}{\pi \hbar }\left(
-C_{z}^{z}+C_{z}^{0}-C_{z}^{x}-C_{z}^{y}\right) ,\;\;\;
\label{wlc}
\end{equation}
where $C\left( \mathbf{r},\mathbf{r}\right)$ are the Cooperon
propagators taken at coinciding coordinates. For completeness, in
Eq.~(\ref{wlc}) we have retained the intravalley Cooperons
$C_{z}^{x,y}$, though they are strongly suppressed by trigonal
warping. Following their suppression, the WL correction is
determined by the intervalley modes $C_{z}^{0}$ and $C_{z}^{z}$
but, in the absence of intervalley scattering, the contributions
of $C_{z}^{0}$ and $C_{z}^{z}$ are equal in magnitude, so that
they cancel. Intervalley scattering due to atomically sharp
scatterers breaks this exact cancellation and partially restores
the WL effect.

A general form of time-reversal-symmetric \cite{t-rev} disorder in
bilayer graphene can be represented following the method used in
the monolayer graphene studies
\cite{mono,CheainovFalkoFO,Aleiner}. That is, we determine
irreducible representations of the symmetry group of the bilayer
crystal formed by $4\times4$ matrices acting in the basis of
states $|\uparrow \rangle _{\mathbf{K}_+,\mathbf{p}}$, $
|\downarrow \rangle _{\mathbf{K}_+,\mathbf{p}}, |\downarrow
\rangle _{\mathbf{K}_-,\mathbf{p}}, |\uparrow \rangle
_{\mathbf{K}_-,\mathbf{p}}$, and, then, use them to construct the
general $t \rightarrow -t$ symmetric disorder
${\hat{u}(\mathbf{r})}$:
\begin{eqnarray}
&&{\hat{u}}=u(\mathbf{r})+\sum \Sigma _{s}\Lambda _{l}u_{sl}
(\mathbf{r});\;\;{s,l=x,y,z};  \label{h1-2} \\
&&\Sigma _{x}=\Pi _{z}\otimes \sigma _{x},\;\Sigma _{y}=\Pi _{z}\otimes
\sigma _{y},\;\Sigma _{z}=\Pi _{0}\otimes \sigma _{z},  \notag \\
&&\Lambda _{x}=\Pi _{x}\otimes \sigma _{z},\;\Lambda _{y}=\Pi _{y}\otimes
\sigma _{z},\;\Lambda _{z}=\Pi _{z}\otimes \sigma _{0}.  \notag
\end{eqnarray}
Here, $\vec{\Sigma}=(\Sigma _{x},\Sigma _{y},\Sigma _{z})$ and
$\vec{\Lambda} =(\Lambda _{x},\Lambda _{y},\Lambda _{z})$ form two
mutually commuting algebras $[\Sigma _{s_{1}},\Sigma
_{s_{2}}]=2i\varepsilon ^{s_{1}s_{2}s_{3}}\Sigma
_{s_{3}},\;[\Lambda _{l_{1}},\Lambda _{l_{2}}]=2i\varepsilon
^{l_{1}l_{2}l_{3}}\Lambda _{l_{3}},$ and $[\vec{
\Sigma},\vec{\Lambda}]=0$. Both $\Sigma_{s}$ and $\Lambda_{l}$
invert sign upon time reversal \cite{t-rev}, whereas their
products remain invariant under $t\rightarrow -t$ transformation.
The first term in $\hat{u}$ represents the potential of remote
charges which is made short range by screening by 2D electrons
$\langle u(\mathbf{r}) u(\mathbf{r^{\prime}}) \rangle \sim u^{2}
\delta (\mathbf{r-r^{\prime}})$. It affects electrons in both
layers equally and manifests itself through the scattering rate
$\tau_{0}^{-1} = \pi \gamma u^{2}/\hbar^{2}$ where $\gamma =
m/(2\pi \hbar ^{2})$ is the density of states per spin in each
valley \cite{socomment}. Disorder $u_{zz}$ generates a random
difference between energies on the $A$ and $\tilde{B}$ sites
(bottom and top layers). The two terms containing $u_{xz}$ and
$u_{yz}$ originate from fluctuations in hopping and they scatter
electrons within each valley, whereas the other terms in $\hat{u}$
take into account intervalley scattering. For simplicity, we
assume that different types of disorder are uncorrelated, $\langle
u_{sl}(\mathbf{r})u_{s^{\prime }l^{\prime }}(\mathbf{r}^{\prime
})\rangle =u_{sl}^{2}\delta _{ss^{\prime }}\delta _{ll^{\prime
}}\delta (\mathbf{r}-\mathbf{r}^{\prime }) $, and, on average,
isotropic in the $x-y$ plane: $u_{xl}^{2}=u_{yl}^{2}\equiv u_{\bot
l}^{2}$, $u_{sx}^{2}=u_{sy}^{2}\equiv u_{s\bot }^{2}$. The
corresponding scattering rates $\tau _{sl}^{-1}=\pi \gamma
u_{sl}^{2}/\hbar $, where $\tau _{sx}^{-1}=\tau _{sy}^{-1}\equiv
\tau _{s\bot }^{-1}$ and $\tau _{xl}^{-1}=\tau _{yl}^{-1}\equiv
\tau _{\bot l}^{-1}$ can be combined into the intervalley
scattering rate $\tau_{\mathrm{i}}^{-1}=4\tau _{\bot \bot
}^{-1}+2\tau _{z\bot }^{-1}$ and the intravalley rate
$\tau_{\mathrm{z}}^{-1}=\tau _{zz}^{-1}$ both of which lead to an
additional suppression of intravalley modes. Intervalley
scattering also leads to the relaxation of $C_{z}^{0}$ although it
does not affect the valley-symmetric mode $C_{z}^{z}$. All the
scattering mechanisms limit the transport time $\tau^{-1} =
\tau_{0}^{-1} + \sum_{sl} \tau_{sl}^{-1}$.

Two low-gap modes $C_{z}^{0}$ and $C_{z}^{z}$ obey the following
equations,
\begin{eqnarray}
\left[ 2 \tau_{\mathrm{i}}^{-1} + \tau_{\varphi
}^{-1}+D\mathbf{\tilde{P}}^{2}-i\omega \right] C_{z}^{0} \left(
\mathbf{r}, \mathbf{r^{\prime }}\right) = \delta \left(
\mathbf{r}-\mathbf{r^{\prime }} \right), \nonumber \\
\left[ \tau _{\varphi }^{-1}+D\mathbf{\tilde{P}}^{2}-i\omega
\right] C_{z}^{z} \left( \mathbf{r}, \mathbf{r^{\prime }}\right) =
\delta \left( \mathbf{r}-\mathbf{r^{\prime }}
\right),\label{lowgap}
\end{eqnarray}
where we included an external magnetic field,
$\mathbf{B}=\mathrm{rot}\mathbf{A}$ in $\nonumber
\mathbf{\tilde{P}}=(i\mathbf{\nabla }+ \tfrac{2e}{c\hbar
}\mathbf{A})$, and inelastic decoherence, $\tau
_{\varphi}^{-1}(T)$.

Equations (\ref{wlc},\ref{lowgap}) yield the zero field WL
correction to the resistivity and the WL MR,
\begin{eqnarray}
\frac{\delta \rho \left( 0\right) }{\rho} &=& \frac{e^{2}\rho}{\pi
h} \ln \left( 1 + 2\frac{\tau_{\varphi }}{\tau_{\mathrm{i}}}
\right) + \delta_0 , \label{MRbi} \\
\frac{\Delta \rho (B)}{\rho} &=& -\frac{e^{2}\rho}{\pi h} \left[
F(\frac{B}{B_{\varphi }}) - F(\frac{B}{B_{\varphi }+
2B_{\mathrm{i}}}) \right] + \delta (B) , \nonumber
\end{eqnarray}
where $B_{\varphi ,\mathrm{i}}= \hbar c/(4De\tau_{\varphi,
\mathrm{i}})$. Equation~(\ref{MRbi}) gives a complete description
of the crossover between two extreme regimes mentioned at the
beginning \cite{socomment}. It also includes small contributions
of the suppressed intravalley Cooperons, $\delta_0 = [2 e^{2}\rho
/(\pi h)] \ln ( \tau_{\varphi} \tau_{\ast} / [\tau (\tau_{\ast} +
\tau_{\varphi})])$ and $\delta (B) = - [2 e^{2}\rho /(\pi h)] F[
B/ (B_{\varphi} + B_{\ast})]$, where $\tau_{\ast}^{-1} =
\tau_{\mathrm{w}}^{-1} + 2\tau_{\mathrm{z}}^{-1} +
\tau_{\mathrm{i}}^{-1}$ and $B_{\ast}= \hbar c/(4De\tau_{\ast})$.
This permits us to account for a possible difference between the
warping time $\tau_{\mathrm{w}}$ and the transport time $\tau$.
According to Eq.~(\ref{MRbi}) WL MR in bilayer graphene sheet
disappears as soon as $\tau_{\mathrm{i}}$ exceeds
$\tau_{\varphi}$, whereas in structures with $\tau_{\varphi} >
\tau_{\mathrm{i}}$, the result Eq.~(\ref{MRbi}) predicts the WL
behaviour, as observed in \cite{exeter}. Such WL MR is saturated
at a magnetic field determined by the intervalley scattering time,
instead of the transport time as in usual conductors, which
provides the possibility to measure $\tau_{\mathrm{i}}$ directly.

Also, it is interesting to consider a small device of bilayer
graphene with both sizes, length $L$ and width $L_{\perp}$, less
than the electron coherence length $\sqrt{D\tau_{\varphi}}$. Since
the edge of graphene scatters electrons between $\mathbf{K_{+}}$
and $\mathbf{K_{-}}$ valleys, in a wire ("ribbon") with $L_{\perp}
< L$ the sample width starts playing the role of the intervalley
scattering length, even though intervalley scattering in the bulk
of the material may be irrelevant, $L_{\mathrm{i}} =
\sqrt{D\tau_{\mathrm{i}}} \gg L$. In this case, in a wire with
conductance $G = G_{\mathrm{classical}} + G_{WL} + \delta G$ both
WL and universal conductance fluctuations (UCFs) \cite{UCF}, would
be described by the usual results established for the orthogonal
symmetry class in disordered systems \cite{UCF}, where UCFs and
the WL parts are of the same order, $G_{WL} \sim
-\frac{2}{3}\frac{e^2}{h} , \quad \langle \delta G^2 \rangle^{1/2}
\sim \sqrt{\frac{8}{15}}\frac{e^2}{h}$. In contrast, in a short
and wide strip with $L_{\perp} > L$ and no intervalley scattering
(either off defects in graphene or contacts), WL in most of the
cross-section is suppressed by warping, whereas the UCFs (which
are immune to such symmetry breaking) are not. Therefore, in a
strip $G_{WL} \sim \frac{e^2}{h} , \quad \langle \delta G^2
\rangle^{1/2} \sim 4 \sqrt{\frac{L_{\perp}}{L} } \frac{e^2}{h} $,
which not only reflects a larger variance of UCFs in broad samples
($L_\bot \gg L$), but also the fact that in a valley-degenerate
system two valleys give coherent contributions towards the
observable conductance fluctuations.

In conclusion, we have shown that $\mathbf{p} \rightarrow -
\mathbf{p}$ asymmetry of the electron dispersion in each valley of
bilayer graphene leads to unusual (for conventional disordered
conductors) behavior of interference effects in electronic
transport. Without intervalley scattering, trigonal warping of the
electron dispersion near the center of each valley leads to a
suppression of weak localization while intervalley scattering
restores it. This behavior is expected in experimentally available
bilayer graphene structures \cite{QHEbi, exeter}, for which
$\tau_{\textrm{w}} \sim \tau$. In contrast, the universal
conductance fluctuations are not reduced by trigonal warping and
may be even stronger in a system without intervalley scattering
than in systems with short $\tau_{\textrm{i}}$: a completely
opposite behavior to the WL effect. Such behavior of bilayer
graphene is similar to that of monolayer graphene \cite{mono,
beenakker}, despite the fact that electrons in these two materials
have different chiralities and can be attributed different Berry
phases: $\pi$ in monolayers, $2\pi$ in bilayers
\cite{Novoselov,McCann}. More generally a suppressed weak
localization magnetoresistance and its sensitivity to intervalley
scattering are specific to all graphitic films independently of
their morphology due to lower (trigonal) symmetry of the
wavevector $\mathbf{K}$ in the corner of the hexagonal Brillouin
zone of a honeycomb lattice crystal. Such a feature of the band
structure does not affect the Josephson proximity effect in
superconductor-graphene devices. This is because the propagation
of a spin-singlet Cooper pair in graphene is related to the
valley-symmetric Cooperon $C_z^z$, whereas Fermi statistics forbid
the appearance of the other long-living mode $C_z^0$ in the
superconducting order parameter.

Authors thank V.~Cheianov, A. Savchenko, and A. Geim for helpful
discussions and authors of \cite{exeter} for experimental details.
This project has been funded by EPSRC grant EP/C511743.


\begin{thebibliography}{99}
\bibitem{socomment} We take into account
double spin degeneracy of carriers.
\bibitem{kpoints} Corners of the hexagonal Brillouin zone are $\mathbf{K}_{\pm }
= \pm ({\textstyle\frac{2}{3}}ha^{-1},0)$ ($a$ is the lattice
constant). In the valleys $\mathbf{K}_{\pm}$, the sublattice
spinor ${|\uparrow \rangle }_{\mathbf{K}_{+}}\equiv [1,0,0,0]$
corresponds to sites $A$, spinor ${|\uparrow \rangle
}_{\mathbf{K}_{-}}=[0,0,1,0]$ to $\tilde{B}$, respectively.
\bibitem{t-rev} In the basis used in this paper time reversal $T(W)$ of an
operator is given by $T(W)=\Pi_{x}\sigma_{x}W^{\ast} \Pi_{x}
\sigma _{x}$.
\bibitem{beenakker} At energies $|\epsilon| \ll \epsilon_L$ (below
the Lifshits transition \cite{McCann}), the bilayer spectrum in
each of the four Fermi surface pockets is linear, and the integral
Berry phase $2\pi$ in bilayer graphene \cite{QHEbi,McCann} is
divided into the Berry phases $\pi$ in each of three side pockets
and $-\pi$ in the central one.
\end{thebibliography}
\end{document}